\documentclass{elsart}
\usepackage{natbib,psfig}

\newcommand{\lsim}{\raise0.3ex\hbox{$<$}\kern-0.75em{\lower0.65ex\hbox{$\sim$}}}
\newcommand{\gsim}{\raise0.3ex\hbox{$>$}\kern-0.75em{\lower0.65ex\hbox{$\sim$}}}

\begin{document}
\runauthor{Wold et al.}
\begin{frontmatter}
\title{The surface density of Extremely Red Objects in high-z quasar fields}
\author[margrethe]{M. Wold}
\author[lee]{L. Armus}
\author[gerry]{G. Neugebauer}
\author[tom]{T.H. Jarrett}
\author[matt]{M.D. Lehnert}

\address[margrethe]{JPL Postdoctoral Researcher, SIRTF Science Center/Caltech}
\address[lee]{SIRTF Science Center/Caltech}
\address[gerry]{Caltech}
\address[tom]{IPAC/Caltech}
\address[matt]{Max-Planck-Institut f{\"u}r Astrophysik}

\begin{abstract}

We report on a study of the surface density of Extremely Red Objects (EROs) in
the fields of 13 radio-loud quasars at $1.8 < z < 3.0$ covering a total area of 
61.7 arcmin$^{2}$. There is a large variation in the ERO surface density
from field to field, and as many as 30--40 \% of the fields have roughly 4--5 times more
EROs than what is expected from a random distribution. The average 
surface density exceeds the value found in large random-field surveys by 
a factor of 2--3, a result which is significant at the $>3\sigma$ level. 
Hence, it appears that the quasar lines of sight are biassed towards
regions of high ERO density. 
This might be caused by clusters
or groups of galaxies physically associated with the quasars. However, an equally
likely possibility is that the observed ERO excess is part of overdensities
in the ERO population along the line of sight to the quasars.
In this case, the non-randomness
of quasar fields with respect to EROs may be explained in terms of gravitational
lensing.

\end{abstract}

\begin{keyword}
galaxies: clusters -- quasars: gravitational lensing
\end{keyword}
\end{frontmatter}

\section{Introduction}

EROs are galaxies with very red optical
to near-infrared colours. Several definitions of EROs are in use, but 
the among the most common are $R-K \geq 5$ and 
$I-K \geq 4$. The colour criterion 
selects $z\approx$ 1--2 early-type galaxies with passively evolving 
stellar populations characterized by strong 4000 {\AA} breaks in their 
spectral energy distributions, but also dusty starbursting galaxies 
which have their UV-flux attenuated by internal dust. 
The first EROs were discovered in the deep near-infrared surveys by
Elston, Rieke \& Rieke \cite{elston88}, and since then there have been several 
surveys aimed at studying their spectroscopic \cite{cimatti02},
photometric \cite{ss00,smail02,mannucci02} and clustering properties 
\cite{daddi00,roche02}.

Here, we have searched for EROs in the fields of $z=1.8$--3.0 quasars.
It has been suggested that the fields of high-redshift radio galaxies and quasars 
often have an excess of EROs \cite{mccarthy92,hr94,cimatti00,thompson00,hall01}. 
E.g.\ Cimatti et al.\ \cite{cimatti00} find an excess of 
$R-K \geq 6$ EROs in 14 radio galaxy and quasars fields at $1.5 < z < 2.0$ and
Hall et al.\ \cite{hall01} find that EROs are about 3 times more numerous in the fields
of $z \approx 1.5$ quasars. Several possibilities for the ERO excess exist,
but the favored explanation 
seems to be that the EROs are early-type galaxies associated with the quasars,
perhaps signifying a forming cluster or a galaxy group at the 
quasar redshift.
Another possibility is that the ERO excess traces foreground clusters or overdensities
in the ERO population which gravitationally magnify
the quasars, boosting them into flux-limited samples. In either case, the ERO excess
may represent clusters in formation at cosmologically significant epochs. 
Care must be taken, however, because the angular distribution of EROs on the
sky is characterized by overdensities and large voids \cite{daddi00,roche02}. 
This causes a strong field-to-field variation in the ERO surface density, 
especially if the fields are small. 

\section{The quasar sample}

The quasar fields presented here are part of a larger ongoing survey to 
study the host galaxies (with the NIRC\footnote{Near Infrared Camera} on Keck~I) 
and environments of radio-loud and
radio-quiet quasars at $1.8 < z < 3.0$. The full sample consists of 40 quasars
from the catalogs by Barthel et al.\ \cite{barthel88} and Hewitt \& Burbidge \cite{hb93},
and are nearly evenly distributed between radio-loud and radio-quiet 
systems with comparable $V$ magnitudes and redshifts. 
In order to study the environments of these quasars, we have obtained 
deep $R$ and $K_{s}$ images with the 
COSMIC\footnote{Carnegie Observatories Spectroscopic Multislit and Imaging
Camera} and the PFIRCAM\footnote{Prime Focus Infrared Camera} 
instruments on the Palomar 200 inch telescope.
The $R$-band imaging is completed, and a subset of 13 radio-loud quasar fields have been 
imaged in $K_{s}$. The ERO samples are drawn from an area of 2.1 $\times$2.1
arcmin$^{2}$ around each quasar (the region of overlap of the $R$ and $K_{s}$
imaging). 
The 5$\sigma$ detection limits in the $R$- and the $K_{s}$-filters are 
$\approx25$ and $\approx20.5$, respectively, and this has allowed us 
to search for EROs in the quasar fields using the $R-K \geq 5$ ERO definition.

\section{Results} 

\begin{figure*}
\psfig{figure=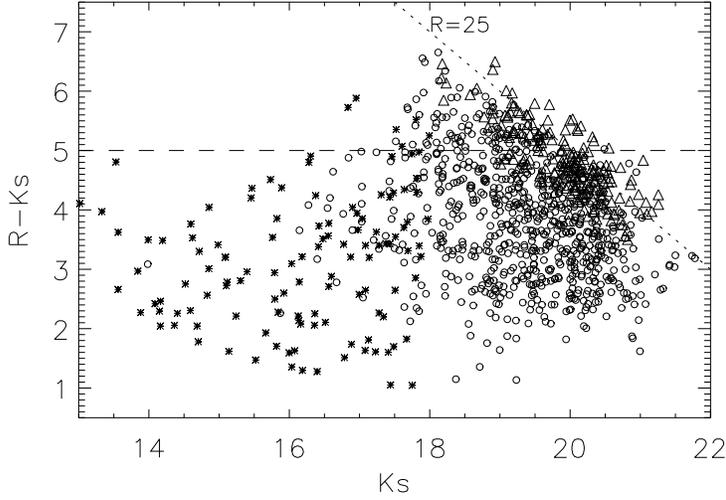}
\caption{Colour-magnitude diagram of objects in the 13 quasar fields. The
open circles are $\geq 5\sigma$ detections in $K_{s}$, and the triangles show 
objects with a detection
significance $< 3\sigma$ in $R$, where a lower limit to their $R-K_{s}$ colour is given
assuming a $3\sigma$ detection in $R$. The asterisks are objects
classified as stars. The horizontal line across the plot shows the ERO definition 
employed in this work.}
\label{figure:fig1}
\end{figure*}

The colour-magnitude diagram of objects 
detected in the quasar fields is shown in Fig.~\ref{figure:fig1}.
In total, there are 160 objects with
$R-K \geq 5.0$. Using the {\sc sextractor} \cite{bertin96} star-galaxy classifier,
and treating all objects at $K_{s} < 18$ with a star-galaxy class greater 
than 0.85 as stars, we find that
six of the 160 EROs are possible stars. Excluding these yields
a surface density of 1.26 EROs per arcmin$^{2}$ at $K_{s} \leq 19$, and 
2.35 arcmin$^{-2}$ at $K_{s}\leq 20$. This surface density
is two--three times higher than what is found in wide-field random surveys,
typically 0.5 at $K_{s} \leq 19$ and 1.4 arcmin$^{-2}$ at $K \leq 20$  
\cite{daddi00,roche02}. Although there is a large variation in the ERO surface
density from field to field, approximately 30--40 \% of the 
quasar fields have notably large overdensities,
a relatively large fraction of the sample in studies like this.

At the resolution of our images 
($\approx1$ arcsec), it is difficult to distinguish between stars and extended objects,
and {\sc sextractor} has problems with star-galaxy separation
at faint magnitudes. We therefore performed the star-galaxy separation
only at $K_{s}<18$.
The contamination of faint red stars among EROs is uncertain, but in our case 
it is likely
to be small since the quasar fields lie at high galactic latitudes ($>35$ deg, 
except three which lie between 21 and 27 deg). 
For completeness, we also carried out
the analysis using star-galaxy separation regardless of $K_{s}$ flux. This resulted
in $\approx20$ \% (as compared to 4 \%) of the EROs being classified as stars.  
This of course lowers the ERO counts, but our conclusions remain unaltered. 
The surface densities using this approach are 1.08 at $K_{s} \leq 19$
and 1.93 arcmin$^{-2}$ at $K_{s}\leq 20$. We note that Mannucci et al.\ \cite{mannucci02}
find $\approx10$ \% stars at $K' < 20$ based on $R$, $J$ and $K'$ imaging.

In Fig.~2, we compare the average cumulative surface density of EROs
in our quasar fields with random field surveys from the literature.
The field counts are represented by the surveys of 
Daddi et al.\ \cite{daddi00}, Roche et al.\ \cite{roche02}
Cimatti et al.\ \cite{cimatti02} and Scodeggio \& Silva \cite{ss00}, and they are
seen to agree well with each other, except the counts in the Chandra Deep Field (CDF)
by Scodeggio \& Silva. This discrepancy is most likely caused by the
inhomogeneous distribution of EROs on the sky,
given that the area of the CDF is only 43 arcmin$^{2}$.
The counts in our quasar fields are seen to be consistently higher than the
random field counts at all flux levels. 
This is not what we expect 
if we are sampling the random field 
ERO population with our 13 widely separated fields. The average ERO surface density
from several small, widely separated fields should in principle be similar to 
the surface density in a large, contiguous field. Hence, it appears that the 
quasar fields are biassed toward regions of high ERO density.

Using the average of Daddi et al's and Roche et al's counts, we 
predict that our quasar fields should contain $\approx27$ EROs at $K_{s} \leq 19$,
whereas the observed number is 79. There is thus a clear 
excess of EROs in the quasar fields,
but given that the EROs cluster strongly and therefore that the field-to-field 
variation is large, is the excess we observe significant? 

In order to answer this question, we calculate (a) the expected 
number of EROs over the total area of our fields from the literature surveys, $N_{field}$, 
and (b), the expected variance, $\sigma^{2}$, of the counts given that the ERO
distribution is non-random, i.e.\ 
$\sigma^{2} = N_{field}\left(1 + N_{field}A_{\omega}C\right)$ \cite{daddi00}, where
$C$ is a factor that depends on the area of the field and $A_{\omega}$
is the amplitude of the ERO angular two-point correlation function as found by Daddi et
al.\ \cite{daddi00} and Roche et al.\ \cite{roche02}.
The significance of the excess in the quasar fields is thereafter
evaluated as $(N_{qso}-N_{field})/\sigma$, where $N_{qso}$ is the observed
number of EROs in the quasar fields. 
The result of this calculation shows that the significance of the
excess at $K_{s} \leq 18$, $19$ and $20$ is 3.1,4.6 and 2.3$\sigma$, respectively.
The ERO excess in the quasar fields therefore appears to be significant,
even after taking into account the large fluctuations in the ERO surface density. 

\begin{figure*}
\psfig{figure=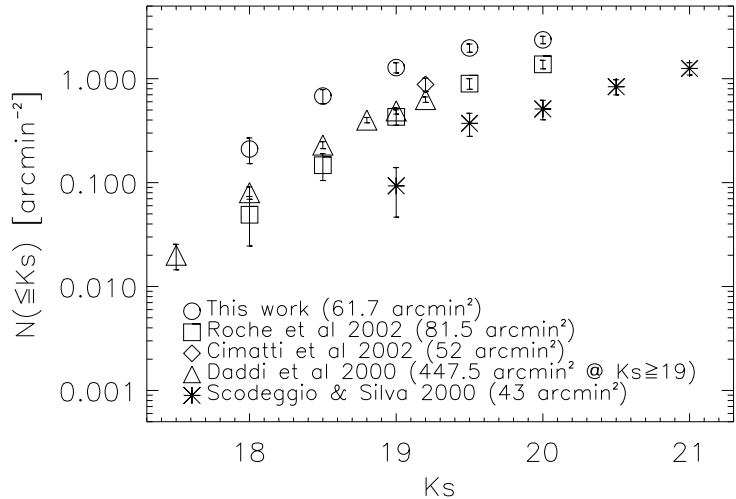}
\caption{Cumulative ERO surface density in our quasar fields (open circles)
compared with various literature samples of EROs in random fields.
Note that a comparison has only been made with surveys defining EROs as
$R-K \geq 5$ or $R-K_{s} \geq 5$. The error bars are Poissonian.}
\label{figure:f2}
\end{figure*}

\section{Discussion}

Our analysis strongly suggests that the quasar lines of sight are biassed
toward regions of high ERO density, and our result is consistent with
that of Cimatti et al.\ \cite{cimatti00} and Hall et al.\ \cite{hall01}. 
At the moment it is impossible to tell whether the EROs are at the same
redshift as the quasars or if they are in the foreground. In order to 
address this, either photometric or spectroscopic redshifts are needed. 
We therefore offer two different explanations for the observed excess and
discuss each of them below: 

The excess might be caused by galaxies, both early types and dusty 
starbursts, at the quasar redshifts. In this case, 
a major part of the EROs could be the early-type population of forming(?) clusters or groups
of galaxies physically associated with the quasars. The evidence in favor
of this consists of observations finding that luminous quasars and
radio galaxies often reside in regions of high galaxy density. 
Another possibility is that the quasars are markers of large-scale structure
traced by the EROs, such as walls or sheets of galaxies, as suggested 
by Thompson, Aftreth \& Soifer \cite{thompson00}. 
Our current $K_{s}$-band images are too small to determine whether the EROs cluster around 
the quasars or if they are just part of a larger-scale overdensity, 
but we are currently awaiting wide-field (9$\times$9 arcmin$^{2}$) 
near-infrared data to properly address this issue. 

There is no evidence for a significant excess 
in the $R$- and $K_{s}$-band counts in our fields, which we might expect if
there are relatively rich clusters in the fields. However, this does not
exclude the possibility that the excess is caused by clusters or groups of 
galaxies because most cluster members would simply be swamped in the
large numbers of foreground sources. 
By selection on $R-K_{s}$ colour we can find the ellipticals at $z\approx$ 1--2
preferentially against much of the confusing foreground galaxies.

Another interesting possibility is that the excess is 
caused by EROs foreground to the quasars, either associated with clusters or 
groups of galaxies, or just overdensities (`clumps') in the ERO population. 
In this case, lensing magnification bias is likely to be important for the 
quasars. By pointing the telescope to a distant quasar, one is more likely to pick 
out lines of sight where there are mass concentrations 
along the line of sight which gravitationally 
amplify the quasar light.

\section{Future and ongoing work}

In order to address the question of how extended the ERO excess is, 
we have started an imaging programme in $J$ and $K_{s}$ using the new 
WIRC\footnote{Wide-field IR Camera} instrument on the Palomar 200 inch.
The WIRC has a field of view of 9$\times$9 arcmin$^{2}$ which matches the size
of our wide-field $R$-band images. We are also planning 
to obtain spectra and morphologies of the EROs.
This will allow us to determine redshifts, spectral types and structural parameters
of the EROs, thereby opening up for an investigation of the Fundamental Plane
of early-type galaxies at cosmologically significant epochs.

\end{document}